\documentclass[apjl]{emulateapj}

\usepackage{amssymb}
\usepackage{epsfig}
\usepackage{longtable} 
\usepackage{amsmath}
\usepackage{amsfonts}
\usepackage{ctable}
\usepackage{graphics}
\usepackage{subfigure}
\usepackage{verbatim}
\usepackage{hyperref}
\newcommand{\lp}{\left (}
\newcommand{\rp}{\right )}

\newcommand{\cf}{$\mathrm{CF}_4$ }
\newcommand{\ccl}{$\mathrm{CCl}_3\mathrm{F}$ }
\newcommand{\nto}{$\mathrm{N_2O}$ }
\newcommand{\ch}{$\mathrm{CH_4}$ }

\begin{document}

\title{Detecting industrial pollution in the atmospheres of earth-like exoplanets}

\author{Henry W. Lin\altaffilmark{1}, Gonzalo Gonzalez Abad\altaffilmark{2}, Abraham Loeb\altaffilmark{2}}
\altaffiltext{1}{Harvard College, Cambridge, MA 02138, USA}
\altaffiltext{2}{Harvard-Smithsonian Center for Astrophysics, 60 Garden St., Cambridge, MA 02138, USA}
\altaffiltext{}{Email: henrylin@college.harvard.edu, ggonzalezabad@cfa.harvard.edu, aloeb@cfa.harvard.edu}





\begin{abstract}
Detecting biosignatures, such as molecular oxygen in combination with a reducing gas, in the atmospheres of transiting exoplanets has been a major focus in the search for alien life. We point out that in addition to these generic indicators, anthropogenic pollution could be used as a novel biosignature for intelligent life. To this end, we identify pollutants in the Earth's atmosphere that have significant absorption features in the spectral range covered by the James Webb Space Telescope (JWST). 
{ We focus on tetrafluoromethane $(\text{CF}_4)$ and trichlorofluoromethane (\ccl\!\!), which are the easiest to detect chlorofluorocarbons (CFCs) produced by anthropogenic activity. We estimate that $\sim 1.2$ days ($\sim 1.7$ days) of total integration time will be sufficient to detect or constrain the concentration of \ccl $(\mathrm{CF_4})$ to $\sim 10$ times current terrestrial level.} 

\end{abstract}

\maketitle

\section{Introduction}
	The search for extraterrestrial intelligence (SETI) has so far been mostly relegated to the detection of electromagnetic radiation emitted by alien civilizations (e.g., \citet{setinature}, \citet{tarter}, \citet{shostak}). These signals could be the byproduct of internal communication \citep{eavesdrop} or perhaps simply the result of the alien civilization's need for artificial lighting \citep{tokyo}.
	
	On the other hand, the search for biosignatures in the atmospheres of transiting Earths has thus far been limited to the search for pre-industrial life. Detecting biosignatures such as molecular oxygen (along with a reducing gas like methane) at terrestrial concentration in the atmospheres of transiting Earths around white dwarfs will be feasible with next-generation technology like the James Webb Space Telescope (JWST) \citep{loebm}. Though molecular oxygen and other signals like the red edge of photosynthesis are strong indicators of biological processes \citep{scalo,seager,kaltenegger} one might ask whether there are specific atmospheric indicators of intelligent life { \citep{camp}} that could be simultaneously searched for.
	
	In this {\it Letter}, we explore industrial pollution as a potential biosignature for intelligent life. This would provide an alternative method for SETI, distinct from the direct detection of electromagnetic emission by alien civilizations. {For definiteness, we defer a brief discussion of Earth-like planets around main sequence stars to \S 3, focusing instead} on Earth-size planets orbiting a white dwarf as \citet{wdhabit} have argued that white dwarfs have long-lived habitable zones, and the similarity in size of the white dwarf and an earth-like planet will give the best contrast between the planet's atmospheric transmission spectrum and the stellar background \citep{loebm}. As in \citet{loebm}, we consider white dwarfs which have been cooling for a few Gyr that have the same surface temperature of our Sun $\sim 6000$ K. { Consequently, the shape of the illuminating spectrum of the white dwarf should resemble that of our Sun, though the normalization should be suppressed by a factor of $\sim 10^4$.} As a result, the habitable zone of a white dwarf is at small orbital radii $\sim 0.01$ AU, increasing the chance of a transit. { The similarity of a $\sim 6000$ K white dwarf spectrum to that of our sun also implies that an earth-like atmosphere is a plausible model of a habitable zone exoplanet's atmosphere as similar photochemistry will take place in both cases \citep{fossati}.} 
		
	As argued in \citet{loebm} and \citet{wdhabit} (and references therein), habitable planets around white dwarfs could plausibly form from a variety of processes including formation out of debris leftover from the expanding red giant and migration from a wider orbit. In fact, observational evidence for habitable-zone exoplanets around white dwarfs is already mounting, with the discovery of very close-in (0.006--0.008 AU) planets around a post-red-giant star \citep{charp}, demonstrating that planets can survive the post main sequence evolution phases of their host star. This result has been interpreted theoretically in \citet{t1, t2, t3, t4, t5, t6}, all giving scenarios by which Earth-analogs could exist in the habitable zone of white dwarfs. In addition, dusty disks have been discovered in a few dozen white dwarfs (see e.g. \cite{zuckerman, dust3, dust1, dust2}), suggesting the existence of exoplanets that can perturb minor planets into the Roche zone of the white dwarf \citep{dust4, dust1}.
	
	On Earth, atmospheric pollution has been carefully studied in the context of global climate change and air quality concerns. It is ironic that high concentrations of molecules with high global warming potential (GWP), the worst-case scenario for Earth's climate, is the optimal scenario for detecting an alien civilization, as GWP increases with stronger infrared absorption and longer atmospheric lifetime. On the other hand, the detection of high concentrations of molecules such as methane $(\text{CH}_4)$ and nitrous oxide $(\mathrm{N_2 O})$ though suggestive, will not be conclusive evidence for industrial pollution, as the contribution from natural sources is comparable to the contribution from anthropogenic sources for the present-day Earth \citep{epa}. Instead, targeting pollutants like CFCs \citep{cfc} is ideal, as they are only produced in significant quantities by anthropogenic activities. In addition, the lifetimes of CFCs vary from $\sim 10^1$ yrs to $\sim 10^5$ yrs. Thus, detection of a short-life CFC will signal an active civilization, whereas the detection of a long-lived CFC will be more probable, as the polluting civilization only needs to exist at any time in the past $\sim 10^5$ yrs.
	
	Our discussion is organized as follows. In \S 2 we calculate the necessary JWST exposure time for detecting different molecules. In \S 3, we focus on CFC-11 and CFC-14. Finally, we summarize our main conclusions in \S 4.
	
\section{JWST exposure time estimate}
	First, we compute the exposure time needed to detect a given molecule in the atmosphere of a planet, given the presence of other molecules and the changing concentration of the molecules with altitude. Although we focus on planets around white dwarfs and JWST, our results are generalizable to other telescopes and planetary systems.
		
	If we are interested in molecule $j$, then the number of signal photons in a given wavelength bin from $\lambda$ to $\lambda +\Delta \lambda$ is given by
		\begin{equation} 
		S_{j,\lambda} = \Delta \lambda \lp \frac{d\mathcal{N}_{wd}}{d\lambda}\rp \epsilon(\lambda) \frac{2\pi}{A_\oplus} \int_{r > r_\oplus} \!\!\!\!\! r dr \, \mathcal{T}_j \lp 1-T_j\rp ,
		\end{equation}
	where $0 \le T_i(r,\lambda) \le 1$ is the intensity transmission coefficient $I/I_0$ of the $i$-th molecule at an altitude of $r-r_\oplus$, $\mathcal{T}_j = \prod_{i \ne j}T_i(r,\lambda)$, and $\epsilon(\lambda)$ is the quantum efficiency of the instrument, which we take to be $0.6$ for the Mid Infrared Instrument (MIRI) on JWST \citep{miri}.
	Further, we assume a blackbody spectrum for a white dwarf with surface temperature of 6000 K:
	\begin{equation}
		\frac{d\mathcal{N}_{wd}}{d\lambda} \propto \frac{\lambda^{-4}}{\exp(2400 \, \text{\AA} /\lambda)-1} 
	\end{equation}
	where $\mathcal{N}_{wd}$ is the number of photons detected from the white dwarf, with the proportionality constant chosen such that at $\lambda = 7000$\,\AA, we have ${d\mathcal{N}_{wd}}/{d\lambda} = 8 \times 10^{-5} \text{s}^{-1} \text{cm}^{-1} \text{\AA}^{-1} \times t_\text{int} \times A_{\text{JWST}},$ where $A_\text{JWST}$ is the effective collecting area of JWST and $t_{int}$ is the total integration time. 
	Here we have assumed that a typical half-occulted white dwarf to be about 18 mag \citep{loebm}. 
	
	To estimate the exposure time needed to detect a given molecule, we use a Monte Carlo approach. We generate a signal $S_{j,\lambda}$, add Poisson noise $N_{j,\lambda} = \sqrt{\Delta \lambda \, d\mathcal{N}_{wd}/d\lambda}$, and then fit a functional form $\sum a_j S_{j}(\lambda)$ to the simulated photon deficit. We take $\Delta \lambda = \lambda/3000$ to simulate medium-resolution spectroscopy that MIRI on JWST will provide and only consider the signal in certain spectral windows to be determined. 
	
	We repeat the process, each time finding best fit parameters $a_j$ and their associated uncertainties. We average these parameters $\bar{a}_j$ over the repeated trials and generate new parameters (averaging again) until all the $\bar{a}_j$ are within $\sim 1\%$ of unity and then take the average uncertainty $s_j$ in $a_j$. Since $a_j = 0$ would indicate no signal, $a_j = 1 \pm 0.2$ corresponds to a $5\sigma$ detection , and in general $a_j = \bar{a}_j \pm s_j$ corresponds to $(\bar{a}_j/s_j) \sigma$ detection. 
	
	Note that the linear fit is approximate; in practice increasing the concentration for $S_{i \ne j}$ will diminish $S_{j}$ signal via the $\mathcal{T}_j$ factor. However the effects of the nonlinearity is small compared to the uncertainty in the molecular concentrations, so we ignore this issue by suppressing the factor $\mathcal{T}_j$ in our estimate. For real surveys where accurate constraints on molecular concentration are desired (and not just an estimate of the necessary integration time), a non-linear fit in concentration should be performed and uncertainties should be estimated by, say, a Markov Chain Monte Carlo method. 
	
	As a check on our calculations, we find that the necessary exposure time for detecting molecular oxygen ($\text{O}_2$) is just $\sim 3$ hrs for an efficiency $\epsilon = 0.15$ and a spectral resolution $R \equiv \lambda/\Delta \lambda = 700$. This is roughly in agreement with the $\sim 5$ hr estimate given by \cite{loebm}, though our method provides a more optimistic integration time as the fitting routine takes into consideration shape information in addition to photon count statistics.
	
%
	
\begin{figure*}
	\centering
 	\includegraphics[height=0.8\textheight, clip=true]{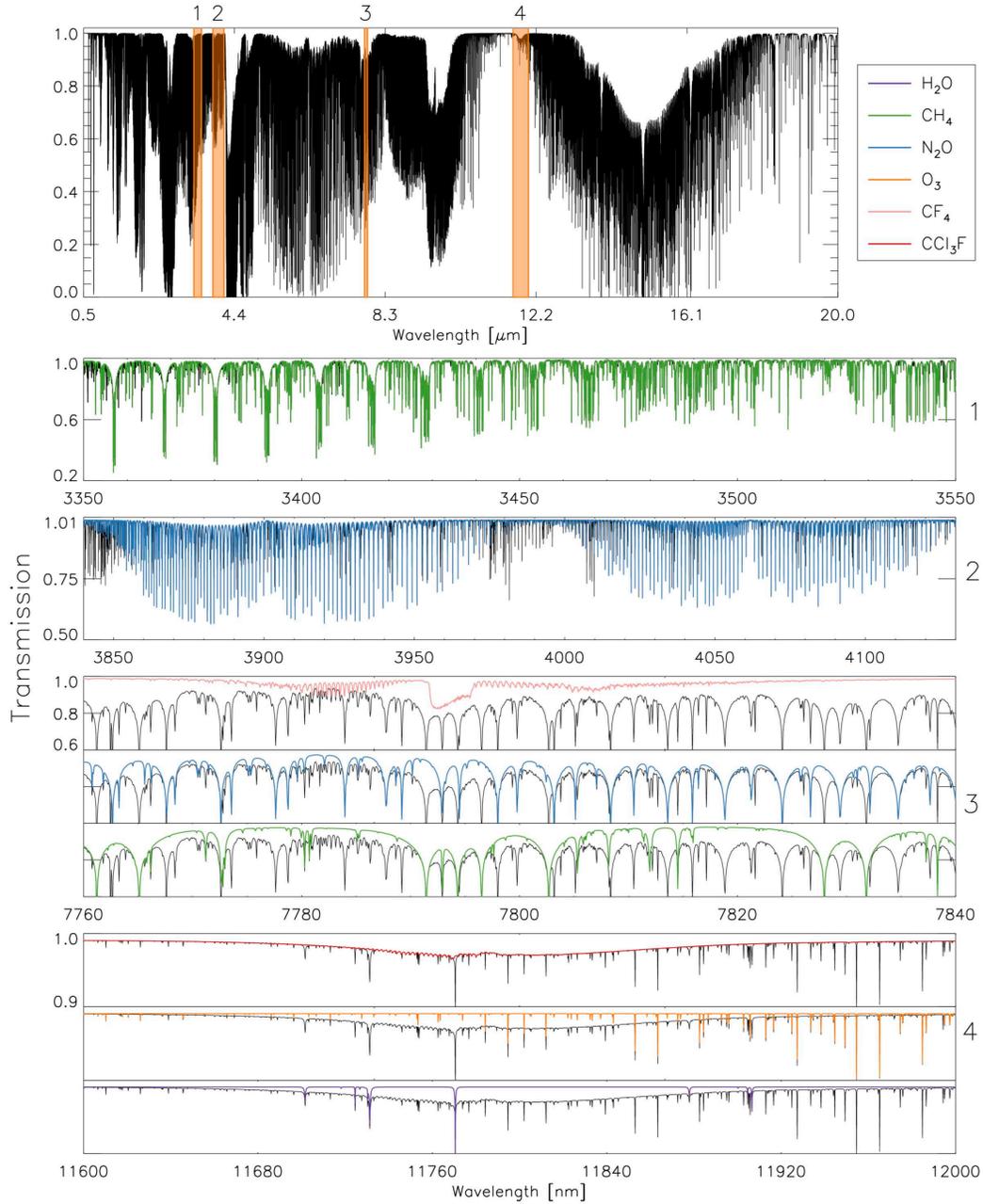}
 	\caption{
	Windows used for detecting \cf and \ccl. The top figure shows the combined transmission $T = I/I_0$ from the most relevant molecules in Earth's atmosphere over the JWST wavelength range. 
	To constrain \cf concentration, we use five intervals in wavelength space where $\mathrm{CF_4}$, $\mathrm{N2O}$, and \ch features dominate. We show in shaded orange three of the five most important regions on this plot (1-3). The fourth region is used to constrain $\mathrm{CCl_3 F}$. The zoomed in versions of these plots are displayed at the bottom.
	\label{comp}
	}
\end{figure*}

\begin{figure*}
	\centering
	\subfigure{
 	\includegraphics[scale=0.4, trim = 0cm 0 0 0, clip=true]{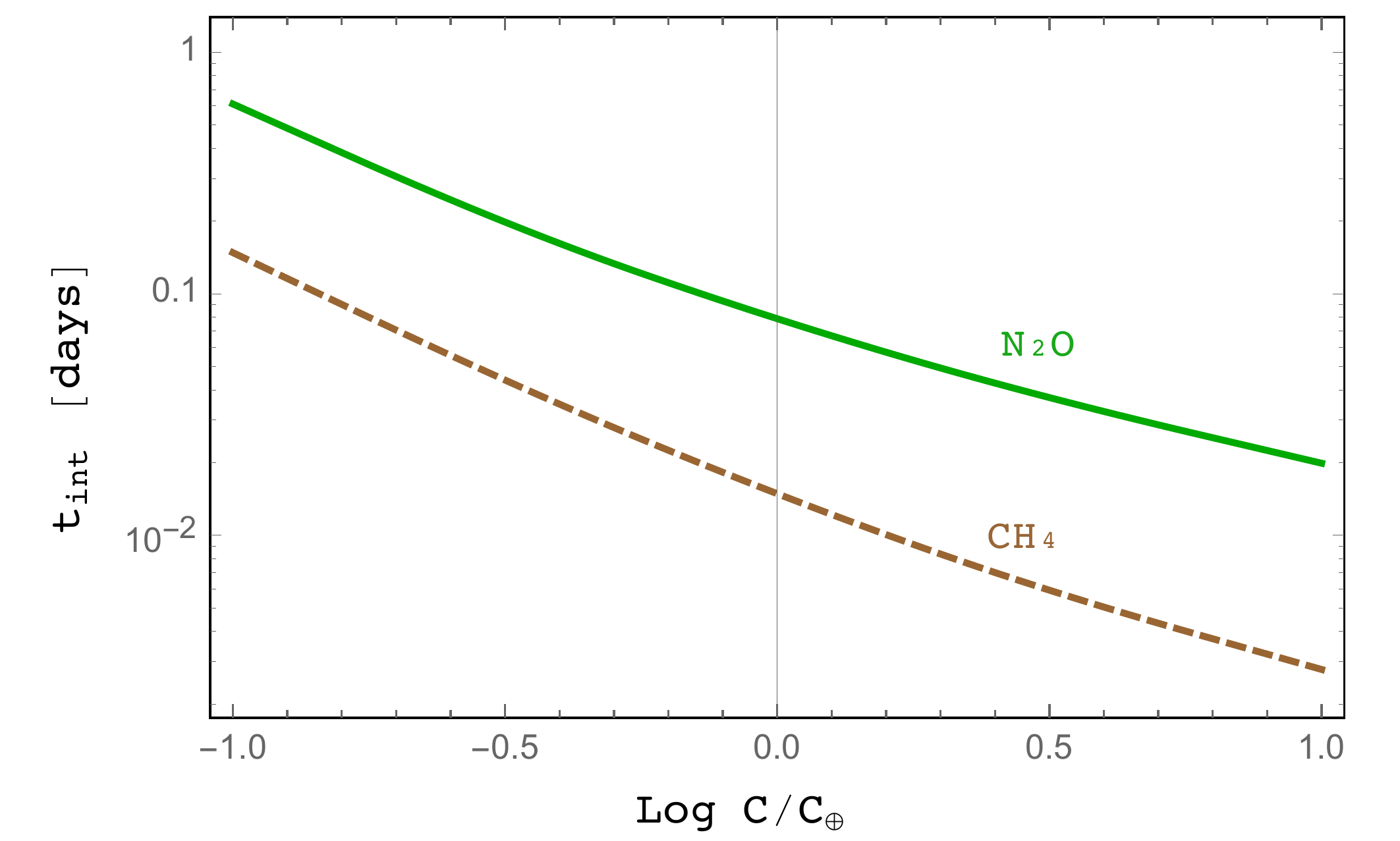}	
	}
	\subfigure{
	\includegraphics[scale=0.4, trim = 0cm 0 0 0, clip=true]{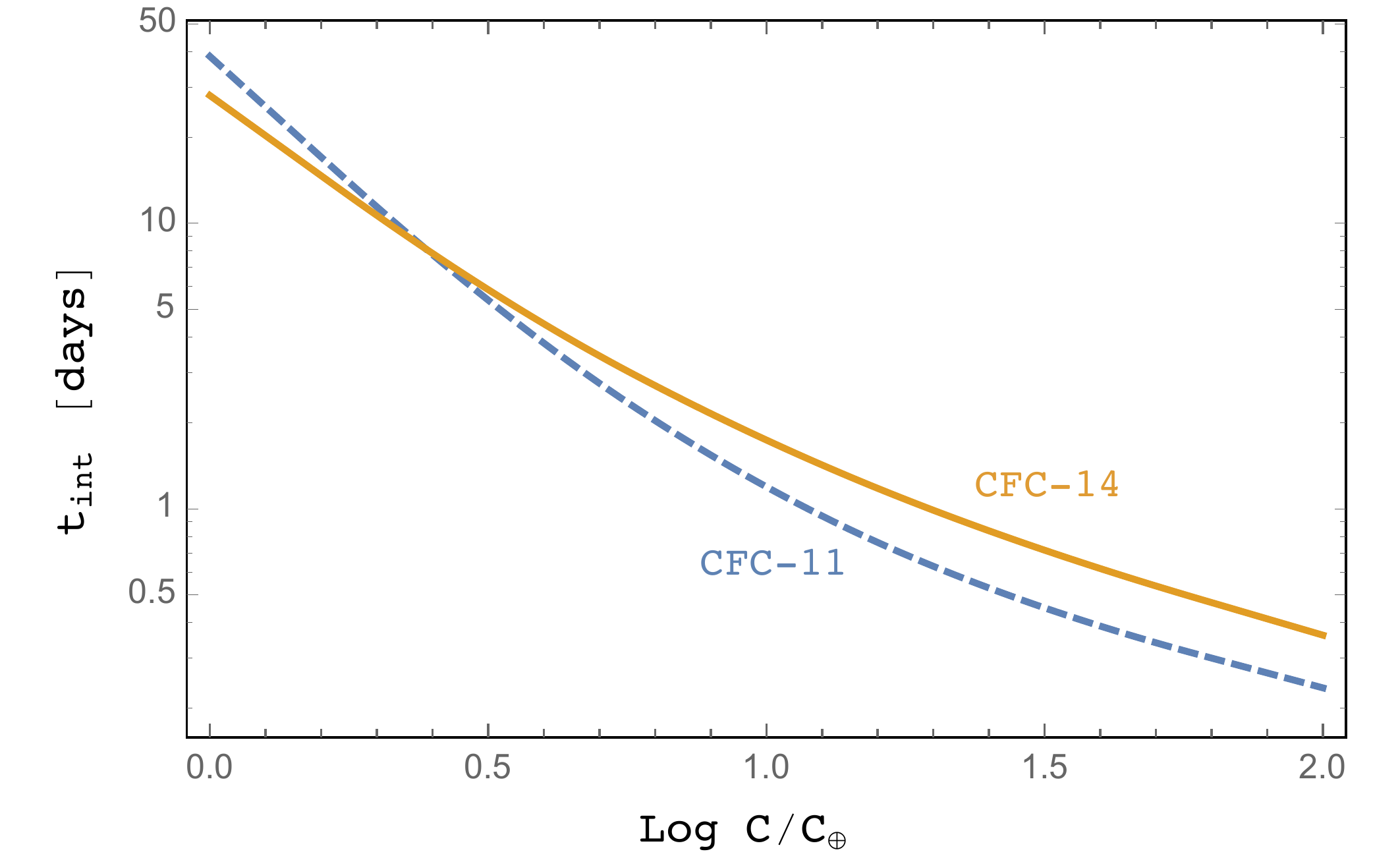}
 	}
	\caption{Exposure time needed to detect various molecules as a function of their normalized concentration $C/C_\oplus$ where $C_\oplus$ is the current terrestrial concentration. On the left panel, \nto and \ch are displayed. On the right, we display \ccl and \cf\!\!. The molecules in the left plot will be considerably easier to detect, requiring $\lesssim 1$ day of observing time for a $5 \sigma$ detection, whereas detection of molecules on the right will be significantly more challenging unless present at concentrations $\sim 10$ times terrestrial levels.
	\label{fig2}
	}
\end{figure*}

\section{Targeting CFC-11 and CFC-14}
	We compile a list of industrial pollutants with small or negligible production from natural sources. Of this list, we identify CFC-14 (\cf \!\!) and CFC-11 (\ccl\!\!) as the strongest candidates for detection. The overlap of the $2\nu_4$ and $\nu_3$ bands around $7.8\, \mu$m is the strongest absorption feature of \cf in the middle infrared region \citep{mcdaniel}. For \ccl the $\nu_4$ band around $11.8 \, \mu$m is the strongest absorption feature  \citep{mcdaniel}. These features are fairly broad, $\sim 0.1\, \mu$m and $\sim 0.4 \, \mu$m for \cf and \ccl respectively.
	The advantage of \cf is that it absorbs at $7.8 \, \mu$m, whereas \ccl absorbs at $\sim 11.7 \,\mu$m. Since these wavelengths are in the Rayleigh-Jeans regime of a 6000 K white dwarf, the photon flux will fall off rapidly with wavelength $\propto 1/\lambda^3$. On the other hand, \ch and \nto are strong absorbers in the regions of interest for $\text{CF}_4$, whereas in the $11$ -- $12 \, \mu$m region where \ccl is strongly absorbing, interference (predominantly from $\text{O}_3$ and $\text{H}_2$O) is less relevant. 
	
	We have calculated the synthetic spectra using the Reference Forward Model\footnote{\url{http://www.atm.ox.ac.uk/RFM/}}. The RFM is a GENLN2 \citep{genln2} based line-by-line radiative transfer model originally developed at Oxford University, under an ESA contract to provide reference spectral calculations for the Michelson Interferometer for Passive Atmospheric Sounding (MIPAS) instrument \citep{mipas}. In our simulations the model is driven by the HITRAN 2012 spectral database \citep{hitran}. We use the US standard atmosphere\footnote{\url{http://www.pdas.com/atmos.html}} to simulate the temperature, pressure and gas profiles of the major absorbers and the IG2 climatology of CFC-11 and CFC-14 \citep{reme} originally developed for MIPAS retrievals.
	
	Since we are interested in \cf and $\mathrm{CCl_3 F}$, we consider the signal in the following spectral windows: 
	\begin{itemize}
	\item{$7760  \text{ nm}< \lambda < 7840 \text{ nm}$ for $\text{CH}_4$, $\mathrm{N_2 O}$, \cf}
	\item{$11600 \text{ nm} < \lambda < 12000 \text{ nm}$ for \ccl}
	\end{itemize}
	Tighter constraints on \ch and \nto in turn improve the constraints on \cf due to \ch and \nto signatures in the wavelengths where \cf features are present. We therefore consider four additional windows:
	\begin{itemize}
	\item{$2190 \text{ nm}< \lambda < 2400\text{ nm}$ for \ch and \nto}
	\item{$3350 \text{ nm} < \lambda < 3550 \text{ nm}$ for \ch }
	\item{$3840  \text{ nm}< \lambda < 4130 \text{ nm}$ for \ch and \nto}
	\item{$4500 \text{ nm} < \lambda < 4600 \text{ nm}$ for \nto }
	\end{itemize}
	Although \ch and \nto absorb strongly at other wavelengths, these windows minimize the interference from other molecules. 
	
	By demanding a given signal-to-noise ratio (say, $S/N = 5$) we can solve for the total integration time $t_\text{int}$ necessary for a $5\sigma$ detection of the given molecule. Figure \ref{fig2} shows these exposure times as a function of the exoplanet's normalized concentration. Note that lower levels of $\text{CH}_4$ and $\text{N}_2$O concentrations on the exoplanet will lower the necessary exposure times for \cf detection, since these molecules provide most of the interference in the $\sim 7.8 \, \mu$m region. For a \cf concentration of $\sim 10$ times that of the Earth, one needs $\sim 1.7$ days of exposure time, whereas $\sim 1.2$ days will be sufficient to detect \ccl at $10$ times terrestrial levels. The long exposure time needed to detect \cf or \ccl in comparison to the time needed to detect molecular oxygen is mostly a consequence of the factor of $\sim 200$ fewer photons per unit wavelength emitted at $\sim 7.8 \, \mu$m compared to $\sim 0.7 \, \mu$m, the location of the molecular oxygen signature. The situation is even worse for \ccl at $\sim 11 \, \mu$m.
	
	Given the cost of long exposure times, we suggest the following observing strategy:
	\begin{itemize}
	\item{Identify Earth-size exoplanets transiting white dwarfs.}
	
	\item{After $\sim 5$ hours of exposure time with JWST, water vapor, molecular oxygen, carbon dioxide and other biosignatures of unintelligent life should be detectable if present at earth-like levels \citep{loebm}.}
	
	\item{While observing for these conventional biosignatures, methane and nitrous oxide should be simultaneously detected, if they exist at terrestrial levels. Extreme levels of methane and nitrous oxide could be preliminary evidence of runaway industrial pollution.}
	
	\item{If biosignatures of unintelligent life are found within the first few hours of exposure time, additional observing time could be used to reduce the uncertainties on the concentration of the discovered biosignatures and search for additional, rarer biosignatures. Methane and $\text{N}_2$O can be ``subtracted out", and constraints on \cf can be obtained. Constraints on \ccl can be simultaneously obtained.}
	
	\item{Direct exoplanet detection experiments which look for reflection or thermal emission from the planet could then be used to push constraints on industrial pollutants to terrestrial levels. At the same time, these experiments could search for less exotic biosignatures like the ``red edge" of chlorophyll \citep{rededge}. {Detecting molecules in exoplanets around main sequence stars with direct detection techniques will be just as feasible as detecting molecules in exoplanets around white dwarfs since the direct detection $S/N$ is roughly independent of the host star's radius $R_*$ whereas for transits $S/N \sim 1/R_*$.}}
	
	\end{itemize}
	
	It is worth noting that a recent study by \cite{cfest} estimate the atmospheric concentration of \cf at $\sim 75$ parts per trillion (ppt), whereas \cf levels were at $\sim 40$ ppt around $\sim 1950$. Assuming a constant production rate $C = C_0 + \gamma t$, we expect as a very crude estimate that in roughly $\sim 1000$ years, the concentration of \cf will reach 10 times its present levels. Coupled with the fact that the half-life of \cf in the atmosphere is $\sim 50,000$ years, it is not inconceivable that an alien civilization which industrialized many millennia ago might have detectable levels of $\text{CF}_4$. {A more optimistic possibility is that the alien civilization is deliberately emitting molecules with high GWP to terraform a planet on the outer edge of the habitable zone, or to keep their planet warm as the white dwarf slowly cools.}
	
\section{Conclusions}
	We have shown that JWST will be able to detect \cf and \ccl signatures in the atmospheres of transiting earths around white dwarfs if these pollutants are found at concentrations at $\sim 10$ times that of terrestrial levels with $\sim 1.7$ and $\sim  1.2$ days of integrated exposure time respectively, though \nto and \ch can be detected at terrestrial concentrations in $1.9$ hrs and $0.4$ hrs respectively. 
	
	Given that conventional rare biosignatures will already take on the order of $\sim 1$ day to detect, constraints on \cf and $\mathrm{CCl_3 F}$, at $\sim 10$ times terrestrial levels, could be obtained at virtually no additional observing costs. Detection of high levels of pollutants like \cf with very long lifetimes without the detection of any shorter-life biosignatures might serve as an additional warning to the ``intelligent" life here on Earth about the risks of industrial pollution.

\section*{Acknowledgments}
This work was supported in part by NSF grant AST-1312034, and the Harvard Origins of Life Initiative.

	\bibliographystyle{mn2e}
	\bibliography{bib}{}

\end{document}